# Software Testing and Code Refactoring: A Survey with Practitioners

Danilo Leandro Lima
Accenture
Recife, PE, Brazil
daniloleandro@gmail.com

Ronnie de Souza Santos
Cape Breton University & CESAR School
Sydney, NS, Canada
ronnie_desouza@cbu.ca

Guilherme Pires Garcia
Agape2IT
Recife, PE, Brazil
guilhermeg@agap2.pt

Sildemir S. da Silva
CESAR School
Recife, PE, Brazil
sss@cesar.school

Cesar França
CESAR School
Recife, PE, Brazil
franssa.cesar.school

Luiz Fernando Capretz
Western University
London, ON, Canada
lcapretz@uwo.ca

*Abstract*—Nowadays, software testing professionals are commonly required to develop coding skills to work on test automation. One essential skill required from those who code is the ability to implement code refactoring, a valued quality aspect of software development; however, software developers usually encounter obstacles in successfully applying this practice. In this scenario, the present study aims to explore how software testing professionals (e.g., software testers, test engineers, test analysts, and software QAs) deal with code refactoring to understand the benefits and limitations of this practice in the context of software testing. We followed the guidelines to conduct surveys in software engineering and applied three sampling techniques, namely convenience sampling, purposive sampling, and snowballing sampling, to collect data from testing professionals. We received answers from 80 individuals reporting their experience refactoring the code of automated tests. We concluded that in the context of software testing, refactoring offers several benefits, such as supporting the maintenance of automated tests and improving the performance of the testing team. However, practitioners might encounter barriers in effectively implementing this practice, in particular, the lack of interest from managers and leaders. Our study raises discussions on the importance of having testing professionals implement refactoring in the code of automated tests, allowing them to improve their coding abilities.

## I. INTRODUCTION

Software quality is defined as the degree to which the software meets the expectation of clients and users and the extent to which it adheres to both specifications of products and processes [1]; this means that software quality includes not only tasks related to the software evaluation (e.g., testing), but also how the whole software development life-cycle adheres to product standards, processes, and procedures (e.g., reviews, code quality, automation, among others) [2].

Software quality is an umbrella, i.e., an overall evaluation process composed of multiple attributes that can attest that the development of a system was successful [3]. Testing is the most common quality assurance strategy applied in software development. Software testing involves appraising the system functionalities to determine that they behave as expected, e.g., searching for failures [4]. Software testing can be performed on several levels (e.g., unit, component, integration, system), using different approaches (e.g., white box, black box, or grey box), and essentially performed in two ways, either manually or using automation [5].

Manual testing requires a great amount of human interaction, as professionals seek to identify failures by interacting directly with the system and observing its behavior. On the other hand, automated testing relies on building and executing code to simulate human interaction and observe the system behavior to identify failures [6]. Currently, automated testing is prioritized in many projects as this technique is presumed to be faster, less repetitive, and provide higher software coverage (e.g., more tests can be performed at the same time) [7]–[9]. However, relying on automated tests can be costly as they require regular maintenance and frequent updating [10].

Code maintenance is a crucial software quality factor; therefore, automated testing demands from software testing professionals a whole set of technical expertise in coding and versioning, in particular, towards refactoring testing code [10], [11]. Refactoring is the process of changing the code without changing the external behavior of the software. Software developers widely apply this practice to improve the quality of the source code [12], [13]. However, professionals can encounter difficulties implementing refactoring in their code [14].

As refactoring is a practice related to code maintenance, it is only natural that testing professionals who work with automation must use this practice in their work, which raises the following research question:

**Research question:** *How do software testing professionals deal with code refactoring when they are working with test automation?*

From this introduction, our study is organized as follows.

In Section II, we discuss existing studies about refactoring. In Section III, we describe how we conducted the survey, while Section IV presents our results. In Section V, we discuss the implications of our study. Finally, Section VI summarizes our contributions.

## II. BACKGROUND

Refactoring promotes improvements in the software by applying changes to its internal components (e.g., the code) while maintaining its observable behavior, i.e., this is a quality strategy that aims to improve code readability, understandability, and maintenance and provide professionals with opportunities to implement re-engineering [12], [13], [15].

Usually, refactoring is a practice associated with the work of developers/programmers, i.e., those who deal more directly with the source code [16]. Developers know that refactoring improves their code; however, several barriers make it difficult to do so in many software projects. These barriers include a lack of resources, tight deadlines, complex changes in the code, high costs, lack of technical knowledge, lack of management, and unavailability of appropriate tools [14].

Currently, with the dynamic software environment resulting from agile development, software testing professionals are required to acquire more programming skills. Therefore, knowing how to code has become one of the primary abilities requested in job advertisements for testing professionals [17]. Software companies require testing professionals to develop a certain level of programming capabilities to successfully build a career in software testing [18], which means that knowing how to refactor code is a necessary technical skill.

Software testing professionals implement code, especially when working with test automation. Knowing how to automate tests is one of the most valued technical skills in the software industry [17], [19], in particular, because of the benefits of test automation, which include the improvement of product quality, high test coverage, reduced testing time, reliability, reusability, reduced human effort, cost reduction, and increased fault detection [20].

However, there are challenges associated with testing automation that sometimes generates problems for testing professionals, such as high costs, unavailability of appropriate tools, inadequate testing structure, insufficient programming knowledge, and constant need for code maintenance [21]. Many of these challenges are similar to or closely related to the barriers that programmers face when refactoring code [14].

## III. METHOD

In this study, we conducted a cross-sectional survey [22] to explore software testing professionals' experience with refactoring the code of automated tests. Our goal is to understand how testing professionals perceive the general benefits and limitations of code refactoring, which are usually associated with the work of software developers (e.g., programmers). To achieve this goal, we surveyed practitioners following three guidelines for conducting surveys in software engineering [23]–[25]. Figure 1 presents an overview of our study, and our methodology steps are presented below.

### A. Using the Literature to Investigate the Industry Practice

Many studies explore the influence of code refactoring on the general quality of software [12], [13], [15]. Usually, the difficulties of conducting code refactoring are discussed considering the work of developers [14]. However, nowadays, we observe in the software industry an increase in the need for testing professionals to improve their programming skills [17], [19], which includes dealing with code refactoring. Therefore, we designed this survey using the findings published in the literature to explore how testing professionals are experiencing code refactoring when working with test automation tasks in the industry.

### B. Instruments

We built our questionnaire based on several published evidence in the literature about how software developers deal with code refactoring [14]–[18], [21]. In particular, we followed the findings discussed by [14] to explore how the general challenges of code refactoring could apply to software testing. Therefore, our questionnaire is mainly based on the results published in the literature that investigated how developers work with code refactoring, including:

- *Testing professionals' knowledge about test automation and code refactoring*: We asked professionals whether they have experience working with test automation or not since this is a valued technical skill in the software industry nowadays, as discussed in [17], [18]. In particular, the perception of code refactoring would be different depending on the experience of participants with programming.

- *Benefits and challenges of refactoring*: we asked participants what are the benefits and challenges of code refactoring based on what was reported in previous studies [14]–[16], [21]. These questions were built to explore the types of refactoring performed in the code of automated tests, the benefits of this practice, and the challenges associated with refactoring testing code regularly.

- *Practitioners profile*: we asked demographic questions about the participants' backgrounds to understand our sample profile. We asked participants to inform their gender, the location of the company where they work, their education level, their experience level (i.e., years of experience working with software development), and whether they have testing certifications.

- *Volunteer participation*: we started the questionnaire by informing practitioners about the goal of our study. In addition, we made sure that they knew their participation was voluntary and anonymous. Therefore, in order to answer the questionnaire, the participants needed to agree to participate under these conditions.

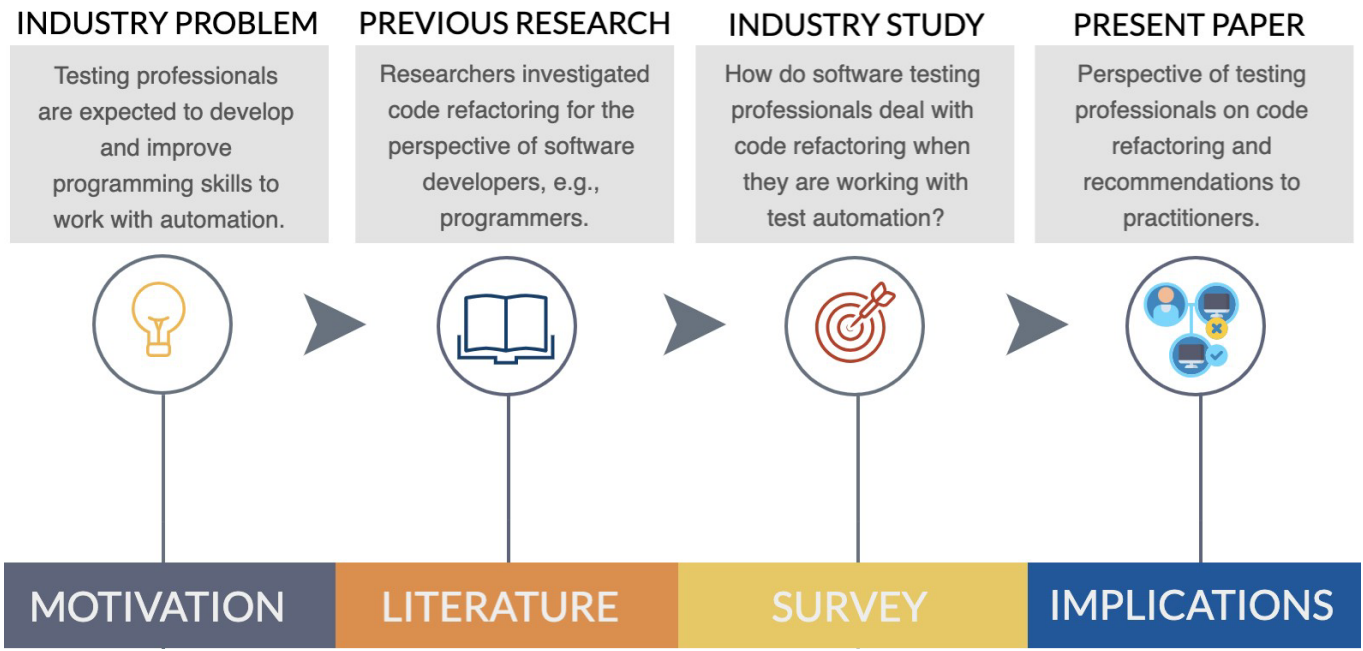

Fig. 1. Study Overview

During the process of designing the survey questionnaire, to increase construct validity and delimit the scope of our investigation, we defined test automation as the automation of test cases performed by software testing professionals at the system level or for regression testing. Therefore, unit testing and integration testing were not explored in this study.

Once all questions were defined, a pilot questionnaire was validated by three senior testing professionals who suggested changes in the wording and terms used in questions 7, 10, 11, and 12 and proposed modifying the sequence of some questions. The participants who validated the pilot questionnaire agreed that presenting the questionnaire with mostly closed-ended questions is a good strategy for exploring the topic since refactoring is not consistently common among testing professionals in the industry. Therefore, having options to be selected would increase the willingness of participants to answer the survey, i.e., extensive open-ended questions or interviews might have increased the difficulty of obtaining data.

The final questionnaire was composed of 12 questions. None of the questions were optional (e.g., participants needed to answer every question). We included an option for the participants to provide other answers different from those listed in the questionnaire, even though, as stated above, our primary goal is to explore in the industry setting a topic that is being mainly explored in the academic context or with different practitioners (e.g., developers). Table I presents the final questionnaire.

*C. Data Collection*

We followed the recommendations for treating software professionals as a hidden population [26], which is a strategy that can mitigate sampling bias in software engineering surveys, as well as support reaching more individuals of a population.

Treating software professionals as hidden populations makes sense in this study because a hidden population cannot be easily defined or enumerated based on existing knowledge [26]. In this case, enumerating all software testing professionals despite their experience with code refactoring is impractical. Recently, the number of studies treating software engineers as a hidden population in the literature is increasing [27]–[29].

Considering this, we applied three techniques to spread our questionnaire and collect data from software testing professionals. All three techniques were non-probability sampling, that is, sampling methods that do not employ randomness. The techniques were applied as follows:

- *Convenience sampling* relied on selecting participants based on availability [26]. Following this technique, we used our extensive network of software testing professionals to share the questionnaire and ask them to answer it, depending on their availability.

- *Purposive sampling* relied on selecting participants from a specific site and inviting them to participate in the study [26]. Following this technique, we contacted the QA manager of a large software company, who forwarded our questionnaire to over 187 software testing professionals in his organization. In addition, we advertised the questionnaire in online software testing communities.

- *Snowballing sampling* relied on identifying some indi-

TABLE I
SURVEY QUESTIONNAIRE

| |
|---|
| 1. We invite you to participate in the survey: Refactoring code of automated tests. This survey is COMPLETELY ANONYMOUS; no information provided can be linked back to you. Answering the questionnaire will take up to 5 minutes. Do you agree to participate?<br>( ) Yes |
| 2. Which gender do you identify with?<br>( ) Female<br>( ) Male<br>( ) Non-binary<br>( ) Prefer not to answer<br><br>3. What is your Country?<br><br>4. What is your highest educational level?<br>( ) High-School<br>( ) Bachelor's degree<br>( ) Post-baccalaureate<br>( ) Master's degree<br>( ) Ph.D.<br><br>5. How long have you been working with software testing?<br><br>6. Do you have any testing certification?<br><br>7. What is your job level?<br>( ) Trainee<br>( ) Junior<br>( ) Mid-level<br>( ) Senior<br>( ) Principal |
| 8. Do you know what code refactoring is?<br>( ) Yes<br>( ) No<br><br>9. Have you ever refactored any test automation code?<br>( ) Yes<br>( ) No<br><br>10. What refactoring techniques do you know or apply? |
| 11. What are the benefits of code refactoring in test automation?<br>( ) Increase of automation reusability<br>( ) Improvement of code readability<br>( ) Improvement of team productivity<br>( ) Improvement of test automation performance<br>( ) Removal of duplicated code<br>( ) No observable benefit<br>( ) Other benefit: ______<br><br>12. What are the challenges of code refactoring in test automation?<br>( ) Changing test code that is working<br>( ) Complexity of changes<br>( ) Lack of knowledge on refactoring<br>( ) Lack of priority<br>( ) No interest from managers/leaders<br>( ) No significant gain<br>( ) No time available (time-consuming)<br>( ) Lack of proper tools<br>( ) Other challenge: ______ |

viduals from the population and asking them to identify other individuals that could participate in the study [26]. Following this technique, we asked participants from the convenience sample to forward our questionnaire to colleagues and co-workers.

### D. Data Analysis

The information collected in this study is mainly quantitative. Therefore, we applied descriptive statistics [30] to analyze the data collected from participants and summarize the information emerging from our data set.

Descriptive statistics allowed us to present the distribution and the frequency of participants' answers regarding their experience refactoring the code of their automated tests and the difficulties faced in completing this task. Since this is a work in progress, more qualitative data will be collected in the upcoming stages of the research.

Qualitative data is not the focus of the survey since we are exploring, in the industrial setting, the benefits and limitations of code refactoring reported in the literature. In particular, we focus on the discussions presented in [11], [14], [16]. In addition, even though we have provided participants with an open-ended option to discuss other aspects related to the theme, the vast majority of our sample stuck with the options and the additional data collected (e.g., other benefits or challenges in addition to the ones listed in the questionnaire) were not statistically representative to be incorporated to the results. Instead, we are keeping them in mind for a future investigation of the theme using qualitative approaches.

### E. Ethics

No personal information about the participants was collected in this study (e.g., name, e-mail, or employer) to maintain participants' anonymity. As mentioned above, we included the beginning of the questionnaire with general information about the study and the research team and asked participants to agree (by checking a yes box) to use their data for scientific purposes.

## IV. FINDINGS

We received 82 answered questionnaires. However, we considered only 80 of them valid, as two participants informed having no experience with software testing; therefore, not being part of our targeted population. Below, we present the results obtained from our descriptive analysis. Table II summarizes the profile of professionals that participated in this survey.

### A. Demographics

In general, our sample is composed of experienced software testing professionals, as 48% of the participants have more than five years of experience working in testing activities, and 46% are working in Senior or Principal positions.

Regarding education, 35% of the participants have a post-baccalaureate certificate in software testing, which shows that over 86% of participants have a high level of training in testing practices.

Further, 26% of our sample are non-male individuals (i.e., women and individuals that preferred not to respond about their gender). Although this is a relatively low rate, previous studies have discussed the lack of diversity in the software industry, which explains why most of our sample is composed of men.

Finally, we have participants working from companies in eight countries. However, almost 59% of our sample is composed of professionals that work for companies in Brazil. Two facts can explain this:

- Two data collection techniques started with Brazilian professionals: convenience and purposive samplings.

- The post-pandemic scenario and the possibility of remote work increased the number of professionals that live in one location and work in another.

Even though most of our sample is from one country, we understand that several participants work on projects that involve international clients, which increases their experience with practices and processes of software development used worldwide.

TABLE II
DEMOGRAPHICS

| | Participants Profile | |
|---|---|---|
| Gender | Male | 59 |
| | Female | 20 |
| | Prefer not to answer | 1 |
| Educational Level | High-School | 4 |
| | Bachelor's degree | 41 |
| | Post-baccalaureate | 28 |
| | Master's degree | 7 |
| Job Level | Trainee | 2 |
| | Junior | 13 |
| | Mid-level | 28 |
| | Senior | 29 |
| | Principal | 8 |
| Experience | 0-1 Years | 5 |
| | 2-4 Years | 13 |
| | 3-5 Years | 23 |
| | 5+ Years | 39 |
| Testing Certification | Yes | 60 |
| | No | 20 |
| Location | Argentina | 4 |
| | Brazil | 53 |
| | Canada | 3 |
| | Germany | 1 |
| | Ireland | 5 |
| | Mexico | 1 |
| | Netherlands | 1 |
| | US | 12 |

### B. Experience with Refactoring Automation Code

We asked participants about their experience with test automation and refactoring, and as a result, 75% (60/80) of our sample reported that they are currently working on test automation activities. This is the same percentage of professionals who performed refactoring in their tests, now, or in previous projects. This means that 25% of our sample (20/80) have never refactored any code.

Further analyzing our data, from the amount of software testing professionals that never refactored any code, 3% (6/20) reported not knowing how to perform the refactoring. However, two of them are currently working with test automation. Most of these professionals are currently at the beginning of their careers (e.g., Junior professionals), which can explain the lack of knowledge about automation and refactoring.

We also asked participants what type of refactoring they used in their automated tests. We compared their answers with the techniques discussed in the literature (e.g., [11], [14], [16]) to build a list of refactoring types. Our results demonstrate that the most used types of refactoring used in testing are:

- *Changing the structure of methods*: to make sure that the automated tests are effectively used by the team, the testing methods are modified to improve readability and understandability.

- *Removing useless code*: to keep up with the project estimations, the automation code is frequently modified by simply commenting on the code, thus, requiring future refactoring to remove dead code.

- *Renaming variables*: to improve code maintainability, readability, and understandability, automation testing refactoring includes renaming variables to more descriptive and concise names.

- *Adding test assertions*: to guarantee test coverage, refactoring the testing code with new assertions is frequently necessary, enabling the verification of new conditions.

- *Other necessary changes*: to guarantee that the automation continues running with no problems, other refactorings might be necessary from time to time, e.g., updating libraries and fixing dependencies related to tools and system versions.

Table III summarizes these results and presents the percentage of answers in our sample.

TABLE III
TYPES OF REFACTORING IN TESTING CODE

| | | % of Participants |
|---|---|---|
| | Changing the structure of methods | 76% |
| | Removing useless code | 75% |
| Refactoring Types | Renaming variables | 50% |
| | Adding test assertions | 49% |
| | Other necessary changes | 5% |

### C. Benefits of Code Refactoring in Test Automation

Based on the general benefits of code refactoring reported in the literature (e.g., [11], [14], [16]) and usually discussed from the perspective of software developers (e.g., programmers and software engineers), we asked software testing professionals how these benefits are perceived in software testing.

As presented in Table IV, our results demonstrate that the general benefits of code refactoring usually observed by software developers also apply to refactoring tasks in test automation, as the testing professionals mostly agreed with what is reported in the literature.

In addition, 35% of our sample (28/80) reported that refactoring improves automation performance, i.e., refactoring the code of automated tests could support these professionals in improving and maintaining the tests, therefore saving time. Moreover, 26% (21/80) reported that refactoring also increases the performance of the testing team, i.e., once the tests are easy to maintain, the professionals would have more time available to focus on other quality activities.

TABLE IV
BENEFITS OF REFACTORING IN TESTING CODE

|  |  | % of Participants |
|---|---|---|
| Benefits | Support automation code maintenance | 75% |
|  | Improvement of code readability | 60% |
|  | Increase of automation reusability | 56% |
|  | Removal of duplicated code | 41% |
|  | Improvement of test automation performance | 35% |
|  | Improvement of team performance | 26% |
|  | No observable benefit | 4% |

### D. Barriers to Code Refactoring in Test Automation

We provided participants with a list of difficulties in conducting refactoring that was previously reported in the literature (e.g., [14]) and asked them to select those that are more challenging when refactoring the code of automated tests. As summarized in Table V, there are at least eight barriers that keep software testing professionals from consistently performing refactoring in their work.

Our result demonstrated that most barriers that challenge testing professionals also challenge software developers. However, software testers struggle with the lack of practical knowledge of refactoring techniques. Testing professionals might not have enough opportunities to deal with refactoring due to how their work is organized and how test automation activities are designed. However, as 75% of our sample have performed code refactoring before (see Section IV-B), this means that this lack of knowledge might refer to complex refactoring changes in the code, i.e., not easy to implement, as pointed out by 20% of the sample.

Almost 80% of our participants reported that refactoring the code of their automated tests would consume too much time; therefore, they need to focus on other activities instead of refactoring. Moreover, 61% of our sample reported a lack of interest from software project managers and team leaders in prioritizing this type of activity in the project. This result indicates that the main barrier faced by software testing professionals is related to how the project activities are planned. Project estimations (e.g., sprint planning) do not usually include code refactoring in testing activities, and this might affect how refactoring is perceived in software testing.

TABLE V
CHALLENGES OF REFACTORING IN TESTING CODE

|  |  | % of Participants |
|---|---|---|
| Challenges | Time-consuming | 79% |
|  | No interest from managers/leaders | 59% |
|  | Lack of knowledge on refactoring | 51% |
|  | Complexity of changes | 20% |
|  | Changing test code that is working | 19% |
|  | Tools unavailability | 16% |
|  | No significant gain | 15% |
|  | Lack of priority | 2% |

## V. DISCUSSION

We start our discussions by comparing our results with the previous literature. Following this, we discuss threats to validity. Finally, we present our plans for future research.

### A. Enfolding the Literature

In the industry, software testing professionals encounter similar difficulties that developers face when dealing with code refactoring, including tight deadlines, complex changes, high costs, lack of specific knowledge, and unavailability of appropriate tools [14]. However, they have an additional challenge to overcome, namely, the lack of interest from managers and leaders in refactoring the code of automated tests.

Although the ability to automate tests is frequently required in job positions and particularly valued in the software industry nowadays [17], testing professionals still need opportunities to practice their programming skills. This would require that software projects include time, resources, tools, and attention to the importance of refactoring for software testing. Software managers should be aware of this barrier and plan project activities accordingly.

This study has implications for research, as there is a lack of studies in the literature that discuss code refactoring of automated tests in the industrial context, e.g., using the experience of professionals to guide the study. We chose to use this perspective, i.e., analyzing how the evidence published in the literature applies to industrial practice as an opportunity to work closely with those effectively experiencing the problem daily. In particular, previous studies have demonstrated the existence of a gap between what academia and practice are interested in software testing [31]. In this study, we tried to bring both worlds together.

In addition, we believe that our findings relate to the discussions raised in recent studies about how software engineering students are not considering careers in software testing because they believe there is a lack of programming activities and challenges in the job [32]. Our results demonstrate that, different from this general belief, software testing involves several challenging activities since refactoring the code of automated tests is a dynamic task performed by testing professionals. Therefore, investing time and resources in refactoring and other coding-related activities in software testing might help increase the attractiveness of this career.

Our results have implications for industry practice as well. Testing is an essential activity in software development, and testing cycles can consume a lot of time and resources in a software project [33]. Therefore, test automation and refactoring are essential practices in software development nowadays. In this sense, our results highlight various benefits of code refactoring to software quality. In particular, because one of the main costs associated with test automation is maintenance, which includes updating the automated tests (i.e., refactoring) [21]. However, to successfully apply this practice, professionals need to understand the barriers presented in this study and identify strategies to overcome them.

*B. Recommendations*

Based on our results, some recommendations can be useful for practitioners and software projects that want to use code refactoring associated with test automation to increase quality:

- Managers and leaders need to plan activities (e.g., sprints) estimating time for refactoring not just for developers but for testing professionals as well.

- Testing professionals need to discuss with their teams the importance of refactoring for test automation, in particular, the relevance of this practice in improving testing performance.

- Developers must be supportive in helping testing professionals to deal with complex refactoring.

- Researchers must focus on the development of tools to support automation in the context of software testing, e.g., automated test code.

In addition, we understand that our results are a starting point for discussions on improving software testing education with training opportunities on testing automation and refactoring. However, since we did not investigate this topic from the perspective of students, we can only hypothesize this.

*C. Threats to Validity*

Since our study targeted software testing professionals, to avoid surveying individuals from a different sample [23], we focused only on professionals who specialized in testing and quality activities on their teams (e.g., software testers, test engineers, test analysts, QAs). We advertised our questionnaire directly to these professionals using convenience and purposive sampling and asked participants to only forward the questionnaire to professionals working in similar roles (snowballing sampling).

To address construct validity [23], [24], our message invitation informed participants that the survey was focused on test automation in the context of system-level testing and regression testing, i.e., unit and integration testing was out of the scope of the research.

For criterion validity and to avoid ambiguity or questions that confuse the participants [23], we validated our pilot questionnaire with three senior testing professionals who supported us in this study; however, they were not included in the sample.

In addition, our survey has a limitation related to how the data was collected. Our questionnaire was mostly based on the benefits and limitations of code refactoring faced by software developers and reported in the literature. In particular, [11], [14], [16] were used to build the list of options provided to participants to select in the survey. Following this, the literature limited our results. We acknowledge that other studies might have been published discussing other general benefits and limitations of refactoring; however, we did not follow a systematic review process to identify the papers in the literature. Therefore, we plan to improve our understanding of code refactoring in software testing by collecting qualitative data from practitioners in the future.

Finally, our survey has a cultural limitation, as our participants are mostly from one country. We plan to solve this limitation by replicating the survey in other regions to increase the participation of professionals from other companies. However, we believe that our current results can be used to inform practitioners about code refactoring in software testing, as 75% of our sample has testing certifications that are recognized worldwide (II), i.e., they are familiar with practices and processes that are applied in software companies from many countries.

*D. Future Work*

Several interesting aspects of the work in software testing can be explored in the industrial context, in particular, considering agile software development and the impact of software quality on software projects, and we expect to investigate them in our future studies.

Specifically, the next steps of this research on supporting test automation will be focused on the following:

- Replicating the survey to increase the sample, targeting testing professionals from other locales to allow generalization of results.

- Conducting focus groups with testing professionals to improve our discussions on the benefits and limitations of refactoring in test automation.

- Exploring tools that can support testing professionals in their refactoring activities;

- Understanding the impacts of the post-pandemic work arrangements (e.g., hybrid work) on quality activities, including test automation and refactoring processes.

We understand that these studies will contribute significantly to the improvement of software quality activities in the industry.

## VI. CONCLUSION

In this study, we investigated the perspective of software testing professionals on test automation and code refactoring. Based on the experience of 80 individuals, we concluded that refactoring offers several benefits, from supporting the maintenance of automated tests to improving the performance of the testing team.

However, practitioners might encounter many difficulties in effectively implementing this practice. Some difficulties are similar to the ones faced by programmers. In contrast, other challenges are observed only in the testing context, such as the lack of attention from managers and leaders toward improving the test automation processes.

Our study has implications for research and practice. We expect that our findings can motivate researchers to develop tools to support testing professionals in coding activities, such as refactoring. As for industrial practice, we are calling the attention of software managers and leaders to the importance of supporting testing professionals in learning, exploring, and applying refactoring in their work to improve the quality of software products.